\documentstyle[prb,aps,tighten,preprint,floats,epsf]{revtex}
\begin{document}

\title{Coherent-Photon-Assisted Cotunneling in a Coulomb Blockade Device}

\author{Karsten Flensberg}

\address{Danish Institute of Fundamental Metrology,
Anker Engelunds Vej 1, DK-2800 Lyngby, Denmark}

\date{27 January 1997}
\maketitle

\begin{abstract}
We study cotunneling in a double junction Coulomb blockade device
under the influence of time dependent potentials. It is shown that
the ac-bias leads to photon assisted cotunneling which in some cases may
dominate the transport. We derive a general non-perturbative
expression for the tunneling current in the presence of
oscillating potentials and give a perturbative
expression for the photon assisted cotunneling
current.
\end{abstract}

\section{Introduction}
\label{sec:intro}

Single electron tunneling (SET) devices have attracted much attention
during the recent years. They can be used to manipulate electrons one by
one, and it is the hope that they can be applied as e.g. memory cells or as
fundamental current standards. However, there are still a number of
unanswered questions regarding the limitations of single electron devices.
One such limitation is the so-called cotunneling process where an electron
charge tunnels through the devices thus giving rise to a leakage current.
This process has been studied extensively both
theoretically\cite{aver89,glaz90matv,aver90,aver94,lafa93} and
experimentally.\cite{eile92,pasq93,paul94,mats95} Experiments have
confirmed the theory of Averin and Nazarov,\cite{aver90} who derived the
cotunneling current for a double junction system at small bias and low
temperatures.\cite{inelastic} It has 
been argued within the present understanding of these
devices that extremely high accuracy could be achieved leading to
metrological applications.\cite{aver89,jens92,fons96} However theory and
experiments differ by several orders of magnitude.\cite{mart94} At present
it not clear what the source of this discrepancy is. The intrinsic noise
properties of the materials due to charge fluctuations\cite{baue93} is one
possible source, but the quantum mechanics of the dynamically driven system
is also an issue which should be considered in order to understand the
fundamental limits.

Therefore an important aspect of single electron devices is their dynamical
property and their response to an applied high frequency signal. So far
theoretical works on single electron circuits\cite{jens92,fons95,fons96}
have used an adiabatic approximation where it is assumed that the static
cotunneling rate can be used at every time step. At first sight this
approximation seems justified when the applied frequency is much lower than
$\Delta/h$, where $\Delta$ is the electrostatic 
energy required to add or remove an electron. As we shall show, 
this is however not so obvious because activated cotunneling is not 
exponentially suppressed but increases as a
power law of applied energy. 
The power law is a consequence of the fact that it is phase-space 
that limits the cotunneling process.

In this paper we investigate the importance of the photon assisted
cotunneling induced by time dependent bias and gate voltages in the
case of a double junction system in the Coulomb blockade regime. Note that
this is different from cases previously discussed in the
literature\cite{odin92,mart93,mart94} where the assisting photons are
supposed to be supplied from an incoherent source, namely the
electromagnetic environment connected to a heat bath.
The coherent-photon-assisted cotunneling is shown to yield an
important and in fact for certain parameters the dominant contribution to
the current. We develop a non-perturbative formalism that combines
cotunneling and sequential tunneling and thus generalizes previous
works which dealt with the static case.\cite{aver94,lafa93,pasq93,koro92}
However, our approach is restricted to the weak coupling limit and
does not include renormalization effects.\cite{scho94}
In the low frequency limit, we separate the photon assisted cotunneling in
two contributions, one corresponding to an adiabatic approximation and
one corresponding to higher order processes. In particular, we derive
a perturbative result for the case of zero dc bias, which we refer to as a
cotunneling electron pump. 

The paper is organized as follows. In Section \ref{sec:model} we introduce
the double junction model. The tunnel current is derived in Section
\ref{sec:scat} using scattering theory and several limits are
considered:weak tunneling, sequential tunneling, the adiabatic limit 
and the case of
small voltage and low temperature where a perturbation expansion is
applicable. In Section \ref{sec:num} numerical results are given and,
finally, conclusions are in Section \ref{sec:concl}. Technical details of
the time dependent scattering theory are in an appendix.

\section{The double junction model}
\label{sec:model}

We use the standard tunneling model for transport through a double barrier
system with Coulomb interactions represented by the electrostatic energies.
The model Hamiltonian reads
\begin{equation}
H = H_0+H_T+H'(t),\label{H}
\end{equation}
where $H_0=H_L+H_R+H_D+H_{\mathrm{int}}$ and $H_{L}$, $H_{R}$, and $H_{D}$
represent the free electron parts in the left lead, in the right lead, and
in the dot, respectively. The charging energy is included in $H_0$ and it
is given by
\begin{equation}
H_{\mathrm{int}} = E_C(N_D-n_g)^2 = E(N_D),
\end{equation}
where $E_C$ is the charging energy, 
$n_g$ is set by the gate voltages and $N_D$ is the number of
electrons in the dot. In Eq.\ (\ref{H}) $H_T$ is the tunneling Hamiltonian
with tunneling matrix element $T_{L(R)}$ connecting the dot to the
left(right) lead. The time dependent parts are contained in $H'$ and we
consider the case where the potentials of the leads and the dot (by the
gate voltage) are modulated in time
\begin{equation}
H'(t) = \sum_{\alpha=L,R,D}\epsilon_\alpha(t) N_\alpha,
\end{equation}
where $N_\alpha$ is the number operator in section $\alpha$.

By a standard unitary transformation (using $\hbar=1$):
$U=\exp\left\{i\int_{-\infty}^{t} dt'\, H(t')\right\},$
we may transform the Hamiltonian into
$H \rightarrow H_0+H_T(t)$, whereby
the time dependence is now present in the tunneling Hamiltonian
\begin{equation}
H_T(t)= \sum_{k,p;\beta=L,R} \Big(T_{\beta}u_\beta(t) d^\dagger_{k}
c_{\beta,p} + \mathrm{H.c.} \Big).
\end{equation}
Here $c$ and $d$ are operators in the leads and in the dot,
respectively, and
\begin{equation}
u_\beta(t) = \exp\left(
i\int_{-\infty}^{t} dt'\, [\epsilon_\beta(t')-\epsilon_D(t')]
\right).
\end{equation}
Without loss of generality we simplify the following discussion by
transferring any non-zero dc part from the time dependent potential to the
static bias or gate voltages. Consequently, if $\epsilon_\alpha(t)$ is
periodic so is $u(t)$ and with the same period.

\section{Scattering theory calculation of the tunneling current}
\label{sec:scat}

In order to calculate the cotunneling rates we generalize the standard time
independent scattering theory and obtain an expression for the average
transition rate $\bar{\Gamma}_{if}$ between an initial and a final state
(both eigenstates of the unperturbed Hamiltonian)
under the influence of a time dependent perturbation (here $H_T(t)$). The
$T$-matrix expansion then forms the basis for a non-perturbative expression
for the current in the limit of small tunneling matrix elements. This
approach thus generalizes previous scattering theory
formulations\cite{aver94,lafa93,pasq93,koro92} to finite frequency and
finite temperature. We furthermore utilize a Breit-Wigner-type 
approximation which does not include renormalization 
effects in the strong tunneling regime.\cite{scho94}

The result for the transition rate is derived in Appendix \ref{app:scat}.
After time averaging it reads
\begin{equation}
\label{Gammageneral}
\bar{\Gamma}_{fi}
= 2\pi \sum_n \delta(E_i-E_f-\omega_n)
\Big|\langle f| T(n)|i\rangle\Big|^2,
\end{equation}
with $\omega_n=2\pi n/T_0$ and $T_0$ being the period of the potentials.
Here $n$ corresponds to the number of photons absorbed. 
The scattering operator $T$ is
\begin{eqnarray}
T(n) &=& H_T(n) + \sum_m H_T(n-m) G_0(m) H_T(m)
\nonumber \\
&&+ \sum_{mm'}H_T(n-m-m') G_0(m+m')
\nonumber \\
&&\times
H_T(m)G_0(m')H_T(m')+\cdots, \label{Tmatrix}
\end{eqnarray}
where the Fourier transforms are defined as
$F(n) = \int dt e^{i\omega_n t} F(t)/T_0$.
and the unperturbed Green's function operator is given by
\begin{equation}
G_0(m) = \frac{1}{E_i+\omega_m  -H_0+i\eta}.
\label{G0}
\end{equation}
The scattering operator $T$ in Eq.\ (\ref{Gammageneral}) can be written as
\begin{equation}
T(n) = \sum_{mm'} H_T(n-m) G(m,m') H_T(m'),
\label{T2}
\end{equation}
where we have defined the Green's function operator by
the Dyson-like equation
\begin{eqnarray}
G(m,m') &=& G_0(m)\delta_{m,m'} \nonumber \\
&& + \sum_{m''} G_0(m)\Sigma(m,m'')G(m'',m'),
\label{Greengeneral}
\end{eqnarray}
where
\begin{equation}
\label{Sigma}
\Sigma(m,m') = \sum_{m''}H_T(m-m'')G_0(m'')H_T(m''-m').
\end{equation}
Note that we have restricted the summation to even powers of $H_T$,
since we shall calculate the transition rate of electron
transfer through two barries.

\subsection{Weak tunneling approximation}
\label{sec:weak}

When the tunneling is not too strong, the off-diagonal element of the
"self-energy" in Eq.\ (\ref{Sigma}) can be neglected, allowing for a
solution of the Green's function. With the notation $A_{\alpha\alpha'}=
\langle \alpha| A | \alpha'\rangle$, the Green's function becomes
\begin{equation}\label{Greenappr}
G_{\alpha\alpha'} (m,m') =
\frac{\delta_{m,m'}\delta_{\alpha,\alpha'}}
{[G_{0,\alpha}(m)]^{-1}+\Sigma_\alpha(m)},
\end{equation}
where
\begin{eqnarray}
\Sigma_\alpha(m) &=& \sum_{\alpha',m''}\langle \alpha |
H_T(m-m'')|\alpha'\rangle
G_{0,\alpha'}(m'')\nonumber\\
&&\times\langle \alpha'|H_T(m''-m')|\alpha\rangle.
\label{Sigmaappr}
\end{eqnarray}
Inserting this into Eq.\ (\ref{T2}), we obtain for the $T$-matrix 
\begin{equation}\label{Tmatrixappr}
T_{if}(n) = \sum_{\alpha,m} \langle i | H_T(n-m)|\alpha\rangle
G_\alpha(m)\langle \alpha| H_T(m)| f\rangle.
\end{equation}
This formula is equivalent to a Breit-Wigner formula 
(see e.g. \onlinecite{landau:qm}),
but here it is generalized to the case of a time dependent perturbation.

\subsection{The tunnel current}
\label{sec:current}

Next we calculate the currents through the double junction system, that is
the transition rate $\gamma^+$ for transferring an electron from left to
right and the rate $\gamma^-$ for electrons moving in the opposite
direction. The total current is $I = e(\gamma^+-\gamma^-).$ In the positive
direction the final state corresponding to an inelastic cotunneling event
is given by
\begin{equation}
|f^+\rangle=  d^\dagger_k c_{L, p} c^\dagger_{R,p'}d_{k'}|i\rangle,
\label{final+}
\end{equation}
whereas $|f^-\rangle$ is obtained by interchanging left and right. The
intermediate state denoted by $\alpha$ in Eq.\ (\ref{Tmatrixappr}), can
take two values, corresponding to an electron being transfered through the
left junction or the right junction first followed by an electron tunneling
through the other junction. We assume that 
the coherence between the individual tunnelings can be neglected, which
means that the tunneling uncertainty time $\hbar/\max(eV,kT,\hbar\omega)$,
is much smaller than $e/I$. In this case, the current can be obtained from the
transition rate between $|f\rangle$ and $|i\rangle$, where $|i\rangle$
is thermally distributed, but with a fixed number of electrons on the
island.

The lifetimes of the intermediate states are given by the imaginary part of
$\Sigma_\alpha$. The real part is neglected since it only gives a small
shift of the electrostatic energy. The lifetime obtained from Eq.\
(\ref{Sigmaappr}) equals the half of the Fermi's Golden Rule 
transition rate for tunneling out of the intermediate state. 
In the present paper, we restrict
ourselves to a two state approximation (denoted by $N=0,1$) and therefore
the lifetime is given by the transition rate from the intermediate state
to the initial state.

By summing over all possible intermediate states while assuming the
electronic occupation in the initial state to be in thermal equilibrium we
finally obtain for tunneling rate in the positive direction
\begin{eqnarray}
\gamma^+ &=& P(0) \int_{-\infty}^\infty
\frac{d\omega_L}{2\pi}\int_{-\infty}^\infty
\frac{d\omega_R}{2\pi}  \nonumber\\
&&\times
\Gamma_L(\omega_L+eV_L)D_{LR}(\omega_L,\omega_R)\Gamma_R(\omega_R-eV_R),
\label{gammap}
\end{eqnarray}
where
\begin{equation}
\Gamma_\alpha(\omega) =\frac{G_\alpha}{e^2} \frac{\omega}{1-\exp(-\beta\omega)}
\equiv G_\alpha f(\omega),
\label{Gamma}
\end{equation}
and where $P(0)$ is the probability that the system initially was in the
state with $N=0$. We must include this factor since our derivation for the
current assumes that the initial state corresponds to the minimum
electrostatic energy.

Above, we have introduced a function $D$, which describes the
propagation of the intermediate states. It is defined by
\begin{eqnarray}
&&D_{LR}(\omega_1,\omega_2)=2\pi\sum_{n}\delta(\omega_1+\omega_2-\omega_n)
\nonumber\\
&&\times\left|\sum_{m}\left(
\frac{u_L(m)u_R^*(n-m)}{\Delta^++\omega_1-\omega_m+
i\Gamma^-(\omega_1-\omega_m)/2}\right.\right.
\nonumber\\
&&\left.\left.+\frac{u_R^*(m)u_L(n-m)}{\Delta^-+\omega_2-\omega_m
+i\Gamma^+(\omega_2-\omega_m)/2}\right)\right|^2
\label{D}.
\end{eqnarray}
Here the first term corresponds to an electron first tunneling through the
left junction followed by an electron tunneling out of the right junction,
while the second term comes from the reverse process. In Eq.\ (\ref{D})
$\omega_m$ is the photon energy in the intermediate state while $\omega_n$
is the energy which is absorbed (emitted) during the cotunneling process. A
non-zero $\omega_n$ thus represents photon assisted cotunneling. The
electrostatic energies in the intermediate states are given by
\begin{equation}
\Delta^\pm = E(\pm1)-E(0).
\end{equation}
The lifetimes in Eq.\ (\ref{D}) are
\begin{equation}
\begin{array}{lcl}
\Gamma^\pm(\omega)& = & \sum_{\alpha = L,R}\Gamma^\pm_\alpha(\omega),\\
\Gamma^\pm_\alpha(\omega)
& = & \sum_n |u_\alpha(n)|^2 \Gamma_\alpha(\omega-\omega_n\mp
eV_\alpha),
\end{array}
\end{equation}
and $\Gamma^\pm$ is thus the transition rate for an electron to tunnel
through one junction in the presence of the oscillating fields. These
rates are well-known from previous works on photon assisted tunneling
in Coulomb blockade systems.\cite{flen92girv,kouw94}
$\Gamma_\alpha$ is the tunneling rate without the ac field given in Eq.\
(\ref{Gamma}).

The reverse rate is obtained by interchanging left and right. We assume
that the bias is such that $V_L=V/2$ and $V_R=-V/2$ (any asymmetry can be
absorbed into the voltage on the island gate). We then obtain the final
expression for the cotunneling current through the device as
\begin{eqnarray}
I &=& \frac{G_LG_R}{e^3} P(0) \int_{-\infty}^\infty
 \frac{d\omega_1}{2\pi}\int_{-\infty}^\infty
\frac{d\omega_2}{2\pi}  \nonumber\\
&&\times \Big[f(\omega_1+eV/2)D_{LR}(\omega_1,\omega_2)f(\omega_2+eV/2)\nonumber\\
&&-f(\omega_1-eV/2)D_{RL}(\omega_1,\omega_2)f(\omega_2-eV/2)\Big].
\label{Ifinal}
\end{eqnarray}
Note that a non-zero current can result even for $V=0$ if the time
dependencies of the left and right junctions are different.
Below we calculate the ``pump'' current based on this mechanism.

\subsubsection{Sequential tunneling limit}
\label{sec:seq}

In the limit of small $\Gamma$ the tunneling current reduces to the
expression for the sequential tunneling described by the familiar master
equation approach.\cite{likh:rev} This is seen as follows.

For small $\Gamma^\pm$, the tunneling rate $\gamma^+$ only has
contributions from the poles of the function $D$. It can be shown that only
diagonal terms of $D$ contributions and $D$ thus reduces to a sum of delta
functions with weights $1/\Gamma^\pm(-\Delta^\mp)$, and in the two state
approximation we keep only the terms corresponding to tunneling between
$N=0$ and $N=1$. After integration the current then becomes equal to
\begin{eqnarray}
&&I_{\mathrm{sequential}} = 
\nonumber\\ &&
P(0) \left[\frac{\Gamma^-_L(-\Delta^+)
\Gamma_R^+(\Delta^+)-\Gamma^-_R(-\Delta^+)
\Gamma_L^+(\Delta^+)}{\Gamma^-(-\Delta^+)}\right]
\end{eqnarray}
and from Fermi's Golden Rule it follows that
$\Gamma^+(\Delta^+)=\Gamma_{01}$ and $\Gamma^-(-\Delta^+)=\Gamma_{10}$,
with $\Gamma_{ij}$ being the rate for tunneling from state $N=i$ to $N=j$
including photon assisted processes.\cite{flen92girv,kouw94} 
From the master equation we have that
$P(0)=\Gamma_{10}/(\Gamma_{10}+\Gamma_{01})$ and the sequential
tunneling result\cite{kouw94} follows.

\subsection{Harmonic potentials}
\label{sec:harmo}

In this paper, we focus on harmonically varying fields. We define
$\epsilon_\alpha(t)= W_\alpha \cos(\omega_0t+\phi_\alpha)$ and find that
the functions $u$ become $u_{(L,R)}(n)= J_{-n}(\alpha_{(L,R)D})$ and
$u^*(n) = [u(-n)]^*$, where $J_n$ are the Bessel functions of the first
kind. The arguments $\alpha_{LD}$ and $\phi_{LD}$ are given by
\begin{mathletters}
\begin{eqnarray}
&&\alpha_{LD}^2
= \alpha_{L}^2+\alpha_D^2-2\alpha_{L}\alpha_D
\cos(\phi_L-\phi_D),\\
&&\alpha_{LD}\sin\phi_{LD} = \alpha_{L}\sin\phi_{L}-\alpha_D
\sin\phi_D,
\end{eqnarray}
\end{mathletters}
where $\alpha_L=W_L/\omega_0$ and $\alpha_D=W_D/\omega_0$ and with similar
relations for $L\rightarrow R$. Using the weak tunneling limit,  where 
the energy arguments of the lifetimes in $D$ are approximated by the pole 
values, we obtain after some manipulations
\widetext
\begin{eqnarray}
D_{LR}(\omega_1,\omega_2)&=&2\pi\sum_{n}
\delta(\omega_1+\omega_2-\omega_n)
\Big|\sum_{m}J_{m-n}(\alpha_{RD})J_m(\alpha_{LD})
e^{im(\phi_{RD}-\phi_{LD})} 
\nonumber\\&&
\times\Big\{ \frac{1}{\Delta^+ +\omega_1-\omega_m+i\Gamma_{10}/2}
+\frac{1}{\Delta^- -\omega_1+\omega_m+i\Gamma_{01}/2}\Big\}
\Big|^2
\label{Dharm}.
\end{eqnarray}
\narrowtext
This is the form which is utilized in the numerical results
reported in Section \ref{sec:num}.
The $D$ function obeys the following relationship
\begin{equation}\label{DLRDRL}
D_{RL}(\omega_1,\omega_2,\Delta^\pm,eV) =
D_{LR}(\omega_2,\omega_1,\Delta^\mp,-eV).
\label{symm} \end{equation}

\subsection{Adiabatic limit}
\label{sec:adia}

In the low frequency limit where the frequency of the  driving fields is
much smaller than the inverse uncertainty time associated with the
cotunneling event, $\omega_0 \ll \Delta/\hbar$,
the frequency dependence in the denominator in the expression for $D_{LR}$,
Eq.\ (\ref{Dharm}), drops out. If it is furthermore assumed that the
$eV,kT, W_{L,R,D} \ll \Delta$, the energy denominators may be replaced by $\Delta^\pm$,
and the summation can be performed. This is equivalent
to using static result by Averin and Nazarov\cite{aver90}
with a time dependent voltage given by
$eV_L(t)-eV_R(t)= eV + W_L\cos(\omega_0
t+\phi_L)-W_R\cos(\omega_0t+\phi_R)$. The current becomes in this
approximation
\begin{eqnarray}
I_{\mathrm{adiabatic}}^{(0)}&=&
\frac{G_LG_R\hbar}{12\pi e^2} V
\Big[(2\pi kT)^2 +(eV)^2 + \frac{3}{2}\tilde{W}^2\Big]\nonumber\\
&&\times
\left(\frac{1}{\Delta^+}+\frac{1}{\Delta^-}\right)^2,
\label{Iadia0}
\end{eqnarray}
where $\tilde{W}^2 = W_L^2+W_R^2-2W_LW_R\cos(\phi_L-\phi_R)$. The two first
terms is the result of ref.\ \onlinecite{aver90}. We see
that the this formula gives no contribution when no dc bias is applied.

In the general case, the adiabatic result can be derived from
the static cotunneling formula, i.e., with $D_{LR}(\omega_1,\omega_2)$
setting $u=1$, and $\omega_0=0$ (which is the formula obtained
by Pasquier {\em et al.}\cite{pasq93}) but allowing for the gate and lead
voltages to vary in time. To next leading order in the applied energies
$eV,kT$  or $W_{L,R,D}$, we obtain after time averaging the formula
\begin{eqnarray}
I_{\mathrm{adiabatic}}^{(1)}
&=&\frac{G_LG_R\hbar}{24\pi e^3}
\big[W_{RD}^2-W_{LD}^2\big]
\nonumber\\ &&\times
\frac{(\Delta^+-\Delta^-)(\Delta^++\Delta^-)^2}{(\Delta^+\Delta^-)^3}
\nonumber\\ &&\times
\big[(2\pi kT)^2+3 \tilde{W}^2/4 \big].
\label{Iadia2}
\end{eqnarray}

In the next section, we calculate the lowest order finite frequency
correction to Eq.\ (\ref{Iadia2}).

\subsection{Perturbative result for photon assisted cotunneling current}
\label{sec:pert}

Here we derive an perturbative result for the cotunneling current in the
case of a ``cotunneling pump'', i.e. the current without a dc bias. We
assume that both the applied frequency and the temperature are low enough
so that we can expand in powers of the energies. In addition, this
approximation has the property that one can neglect the broadening effects
due to the small finite lifetimes, because it is assumed that the energy
is so low that the real transitions (corresponding to zeroes of
the real part of the denominators), do not play a role.

In the case of zero dc bias the cotunneling current is given by Eq.
(\ref{Ifinal}) with $V=0$. Now by using Eq.\ (\ref{symm}) and by
interchanging $\omega_1$ and $\omega_2$ in Eq.\ (\ref{Ifinal}) the
difference $D_{LR}-D_{RL}$ can be written as
$D_{LR}(\Delta^+,\Delta^-)-D_{LR}(\Delta^-,\Delta^+)$. Therefore the pump
current vanishes when the gate voltage is set to zero. This is expected
because in that case there is no preferred tunneling direction.

Next we expand in powers of the applied frequency $\omega_0$ whereby the
integral in Eq.\ (\ref{Ifinal}) can be performed analytically. After
summation over frequencies we obtain the final expression for the pump
current (reinserting $\hbar$)
\begin{eqnarray}
I_{\mathrm{PACT,pump}} &=& \big[W_{RD}^2-W_{LD}^2\big]
\frac{G_LG_R\hbar}{24\pi e^3} 
\nonumber\\&&\times
\frac{(\Delta^+-\Delta^-)(\Delta^++\Delta^-)^2}{(\Delta^+\Delta^-)^3}
\nonumber\\&&\times
\big[(2\pi kT)^2 +3 \tilde{W}^2/4+(\hbar\omega_0)^2\big].
\label{Ipumpfinal}
\end{eqnarray}
The two first terms is the adiabatic result derived in the previous section.
The last term is the lowest order quantum correction to the adiabatic
limit.
\begin{figure}
\vbox to 15cm {\vss\hbox to 15cm
 {\hss\
   {\includegraphics{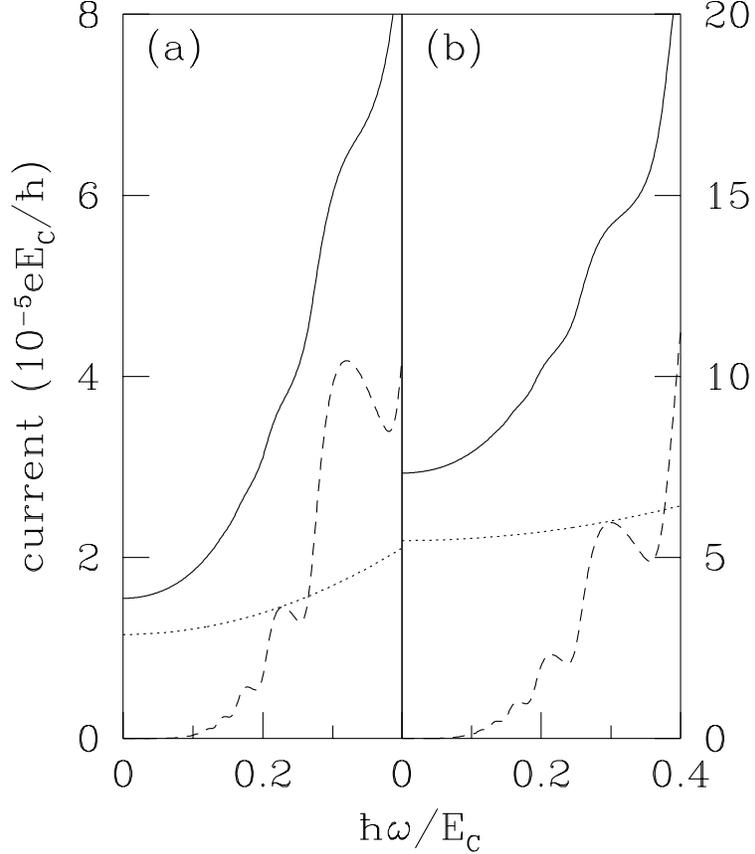}
   }
  \hss}
}
\caption{
The photon assisted cotunneling compared to the
sequential tunneling current (dashed curve)
and the perturbative result (dotted curve) derived
in Section \protect\ref{sec:pert} in the case where only
the left electrode potential is modulated.
The parameters are $W_L$ = 0.5 $E_C$, $W_R=W_D=0$,
$G_L=G_R=$ 0.1 $e^2/h$, $kT= 0.01\, E_C$, and $\Delta^+-\Delta^-=
0.2\, E_C$. Using a typical value for the charging energy $E_C = .4$ meV
corresponds to $kT= 50$ mK and the frequency is scanned up to 40 GHz
in the figure. In (a) there is no applied voltage whereas in (b)
$eV =0.1\, E_C$. }\label{fig:freq}
\end{figure}

\section{Numerical Results}
\label{sec:num}

In the following, we compare expression (\ref{Ipumpfinal}) with the current
through a Coulomb island calculated on the basis of standard master
equation approach\cite{likh:rev}, the socalled sequential tunneling limit
obtained, see Section \ref{sec:seq}. In this calculation we must use the
photon assisted sequential tunneling rates.\cite{flen92girv,kouw94} In
Fig.\ \ref{fig:freq} we show the different current contributions as a
function of frequency for the case where an ac current is applied to the
left lead only. At low frequencies the photon assisted cotunneling is much
larger than the sequential tunneling current. Even with a finite bias
voltage (b) this still holds.

\section{Summary and conclusions}
\label{sec:concl}

In summary, we have shown that the coherent photon assisted cotunneling is
an important contribution to the charge transfer processes in single
electron devices. It must therefore be taken into account when these
devices are operated under periodic time dependent conditions, which is
indeed the case in many potential applications. The photon assisted
contribution that we have pointed out in this paper has not been included
previously in models for SET systems, therefore it
would be interesting to study experimentally the photon assisted cotunneling
process, e.g., by studying a double junction device under the pump
conditions, {\it i.e.} with zero dc bias but asymmetrically biased. 

However, since the coherent-photon-assisted 
cotunneling process is important in a small parameter space only,
it may be difficult to separate the competing tunneling mechanisms
from the process studied here. 
In order to observe the photon assisted cotunneling process one should thus
bias the device as asymmetrically as possible because it is proportional 
to the  {\em difference} between the ac amplitudes, as can be seen in the 
perturbative result in Eq.\ (\ref{Ipumpfinal}), 
whereas the usual photon assisted sequential tunneling 
is additive in the applied ac power. Furthermore, since the cotunneling current
stems from a higher order tunneling processes the resistances should not be too
small. In the numerical examples presented in Fig.\ \ref{fig:freq},
resistances of 0.1 $h/e^2$ were used. In this case the photon assisted 
cotunneling current dominates for small frequencies but it is important to note
that a part of this is due to the adiabatic cotunneling current, {\it i.e.} 
it can be explained by a time averaged dc cotunneling, giving rise
to non-zero time average due to the non-linear bias dependence.
Since the adiabatic contribution is frequency independent it can
in principle be measured in the small frequency limit.
 
The non-perturbatively expression derived in the this paper generalizes
previously derived cotunneling results for the dc 
case\cite{aver94,lafa93,pasq93,koro92} and as in those theories
it does not take renormalization effects into account. Such effects are 
important in the strong tunneling regime\cite{scho94} and a study of these
effects in the time domain is a interesting subject which deserves further 
work.

\section*{Acknowledgment}

Henrik Bruus, Ben Hu, Antti-Pekka Jauho, Hans Dalsgaard Jensen, John Martinis, Dick
Kautz and Mark Keller are acknowledged for useful discussions and comments.

\appendix
\section{Time dependent scattering theory}
\label{app:scat}

\widetext

In this appendix, the standard $T$-matrix approach is generalized
to the case of a periodic time dependent perturbation, $V(t)$,
which is assumed to be turned on adiabatically in the distant past,
i.e. it includes a factor $\exp(\eta t)$ where $\eta=0^+$.
We want to calculate the probability to find the system at time
$t$ in the state $|f\rangle$ given that it started in the initial
state $|i\rangle$, which is given by the square modulus of the
of the overlap, $a_{if}=\langle f|i(t)\rangle$.

First consider the time evolution of an initial state $|i\rangle$
\begin{equation}
|i(t)\rangle = U(t,t_0)|i(t_0)\rangle,
\end{equation}
where $U(t,t_0)$  is the time evolution operator. Expanding  in powers
of $V(t)$ the $n$th order term becomes (for $t_0\rightarrow -\infty$)
\begin{equation}
|i^{(n)}(t)\rangle = (-i)^ne^{-iH_0t}
\int_{-\infty}^{t}dt_1 \int_{-\infty}^{t_1}
\cdots \int_{-\infty}^{t_{n-1}}dt_n\,V_I(t_1)V_I(t_2)\cdots
V_I(t_n)|i\rangle,
\label{nterm}
\end{equation}
where $V_I(t)$ is in the interaction picture. Now define the Fourier
transforms $V(t)= \sum_{n}e^{-i\omega_0 n t}V(n)$ and insert these in
(\ref{nterm}) which allows us to perform the integrations and obtain
\begin{eqnarray}
&&|i^{(n)}(t)\rangle = e^{-iE_i t+\eta t}\sum_{\{m\}}
\exp\left[-i\omega_0 (m_1+\cdots+m_n)t\right]
\nonumber \\
&&\times
G(m_1+\cdots+ m_n)
V(m_1) \cdots G(m_{n-1}+m_n)V(m_{n-1})G_0(m_n)V(m_n)|i\rangle
\end{eqnarray}
where $G_0$ is given by Eq.\ (\ref{G0}) and $\omega_0=2\pi/T_0$.

Taking the square modulus of the overlap $a_{if}$, we obtain
an expression of the form (emphasizing only the time dependence)
\begin{equation}
|a_{if}(t)|^2= \sum_{n \ell}  \sum_{\{m\}\{m'\}}
\exp(-i\omega_0[m_1+\cdots+m_n+m_1'+\cdots+m_\ell']t)e^{2\eta t}\cdots
\end{equation}
The time averaged overlap is found by integrating over one period assuming
that $T_0 \ll 1/\eta$. (This approximation implies that the frequency
of the applied ac signal is much larger than the frequency of the tunnel
events.) Averaging gives that the sum of the two sets of intermediate
photon energies must vanish; below we denote the sum of one set
as $j=m_1+\cdots + m_\ell$. The tunneling rate is obtained as
the time derivative of $|a|^2$ and we then obtain for the
average rate
\begin{eqnarray}
&&\bar{\Gamma}_{if} =  \sum_{j} \frac{2\eta e^{2\eta t}}
{(E_i-E_f-\omega_0 j)^2+\eta^2}
\left|\langle f| V(j) + \sum_m V(j-m)G_0(m)V(m)\right.\nonumber\\
&&\left. + \sum_{m_1m_2}V(j-m_1-m_2)
G_0(m_1+m_2)V(m_2)G_0(m_1)V(m_1)\cdots|i\rangle\right|^2.
\end{eqnarray}
In the limit $\eta \rightarrow 0$, we obtain the result quoted in
the main text.

\narrowtext

\begin{thebibliography}{10}
\bibitem{inelastic}
Here we consider only the inelastic cotunneling process which is expected to be
  dominant for realistic device parameters. See ref.\
  \protect\onlinecite{aver90} for a discussion.
\end{thebibliography}

\end{document}